\documentclass[showpacs,prl,onecolumn,aps,superscriptaddress,preprintnumbers,nofootinbib]{revtex4}
\usepackage[T1]{fontenc}
\usepackage[latin9]{inputenc}
\usepackage{amsmath,amssymb}
\usepackage{epsfig}
\usepackage{bm} 
\usepackage{graphicx}
\usepackage{mathrsfs}
\usepackage{amsmath}
\usepackage{amsfonts}
\usepackage{epstopdf}
\usepackage{color}
\usepackage{pifont}
\usepackage{relsize}
\usepackage{mathtools}
\def\slashchar#1{\setbox0=\hbox{$#1$}     		
   \dimen0=\wd0                                 	
   \setbox1=\hbox{/} \dimen1=\wd1               	
   \ifdim\dimen0>\dimen1                        	
      \rlap{\hbox to \dimen0{\hfil/\hfil}}      	
      #1                                        	
   \else                                        	
      \rlap{\hbox to \dimen1{\hfil$#1$\hfil}}   	
      /                                         	
   \fi}

\renewcommand{\vec}{\boldsymbol}
\newcommand{\beq}{\begin{equation}}
\newcommand{\eeq}{\end{equation}}
\newcommand{\bea}{\begin{eqnarray}}
\newcommand{\eea}{\end{eqnarray}}
\newcommand{\ba}{\begin{array}}
\newcommand{\ea}{\end{array}}

\def\eq#1{{Eq.~(\ref{#1})}}
\def\fig#1{{Fig.~\ref{#1}}}
\newcommand{\bas}{\bar{\alpha}_S}

\newcommand{\nn}{\nonumber}
\newcommand{\bg}{ \bar{\gamma}}

\newcommand{\Lb}{\left(}
\newcommand{\Rb}{\right)}
\newcommand{\h}{\frac{1}{2}}

\newcommand{\pom}{I\!\!P}

\newcommand{\intl}{\int\limits}
\begin{document}

\title{ Summing large Pomeron loops in the saturation region:  dipole-nucleus collision beyond  nonlinear equations.}

\author{Eugene Levin}
\email{leving@tauex.tau.ac.il}
\affiliation{Department of Particle Physics, Tel Aviv University, Tel Aviv 69978, Israel}

\date{\today}


\begin{abstract}
In this paper we found  the dipole-nucleus scattering amplitude at high energies by summing  large Pomeron loops. It turns out that the energy dependence of this amplitude  is the same as for dipole-dipole scattering. It means that  the Balitsky-Kovchegov (BK) equation, which  has been derived to describe this scattering,  can be trusted only in the limited range of  energies: $z \,\leq\, 2\sqrt{\,\kappa\,c}A^{1/6} $

 \end{abstract}
\maketitle

\vspace{-0.5cm}
\tableofcontents

\section{Introduction}
 
\begin{figure}[ht]
    \centering
  \leavevmode
  \begin{tabular} {c c}
      \includegraphics[width=6cm]{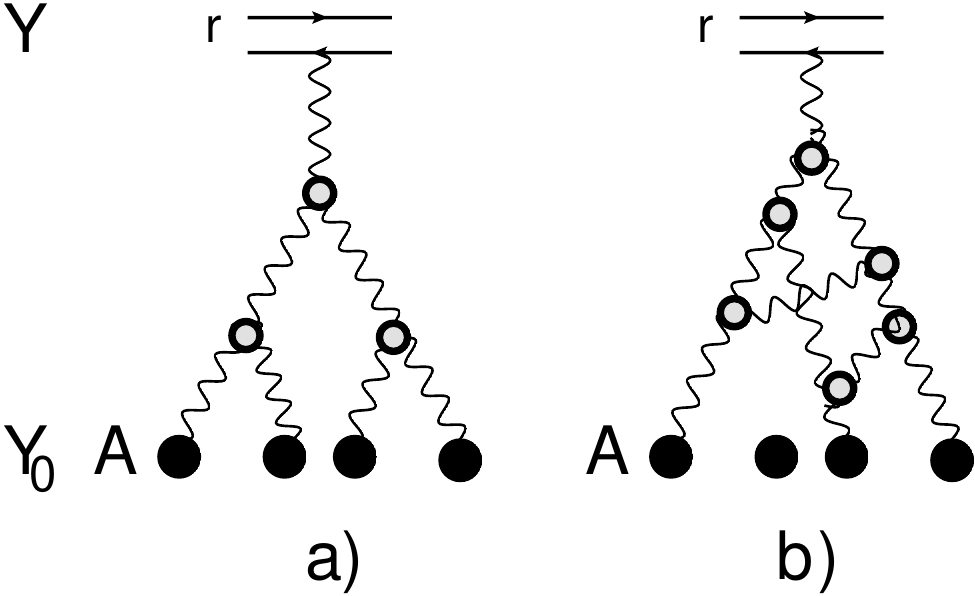}&   \includegraphics[width=9cm]{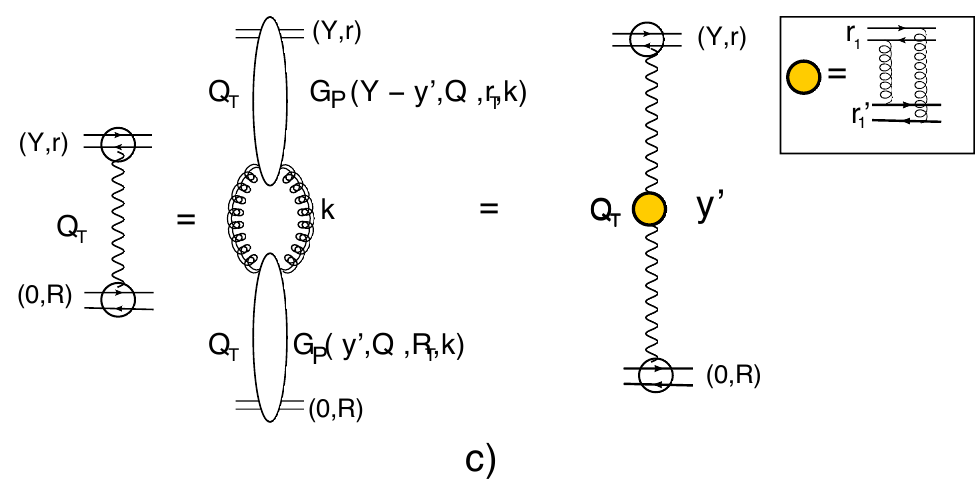}\\
      \end{tabular} 
      
      \caption{ \fig{gen}-a: The  example of `fan'  BFKL Pomeron diagrams that lead to non-linear BK equation.  \fig{gen}-b: The example of the Pomeron loop diagrams ( enhanced diagrams) that we are summing in this paper at very high energy.\fig{gen}-c: $t$-channel unitarity for the BFKL Pomeron. The double helix lines denote the reggeizied gluons. The wavy lines denote the  BFKL Pomeron exchanges.  The black circles  denote the amplitude $\gamma^{BA}$ for interaction of the dipole with the nucleon which we replaced by the dipoles of the size $r'$(see \fig{mpsi}). The gray circles stand for the triple Pomeron vertex. }
\label{gen} 

   \end{figure}

 The goal of this paper is to  sum the large BFKL Pomeron loops\cite{BFKL}\footnote{BFKL stands for Balitsky, Fadin,Kuraev and Lipatov.} for dipole-nucleus scattering at high energies. The scattering amplitude of this process satisfies the nonlinear equation which has been derived in momentum\cite{GLR,MUQI}  and coordinate \cite{MUDI,B,K} representation using different technique\cite{BART,BRAUN,BRN,MV, JIMWLK1,JIMWLK2, JIMWLK3,JIMWLK4, JIMWLK5,JIMWLK6,JIMWLK7, JIMWLK8, LELU}. However, recently it has been shown\cite{KLLL1,KLLL2} that this nonlinear BK\footnote{BK stands for Balitsky and Kovchegov} equation 
, that governs the dilute-dense  dipole  scattering (deep inelastic scattering (DIS) of electron with  nuclei), has to be modified due to  contributions of  Pomeron loops.  Indeed, in the BFKL Pomeron calculus the BK equation stems from summing the `fan'  Pomeron diagrams of \fig{gen}-a \cite{GLR,BRAUN}. The enhanced Pomeron diagrams with the Pomeron loops (see \fig{gen}-b) certainly lead to an additional contribution which we are going to estimate in this paper at ultra high energies. It is well known that 
 in spite of intensive work 
\cite{BFKL,KOLEB,MUSA,LETU,LELU1,LELU,LIP,KO1,LE11,RS,KLremark2,SHXI,KOLEV,nestor,LEPRI,LMM,LEM,MUT,MUPE,IIML,LIREV,LIFT,GLR,GLR1,MUQI,MUDI,Salam,NAPE,BART,BKP,MV, KOLE,BRN,BRAUN,B,K,KOLU,JIMWLK1,JIMWLK2,JIMWLK3, JIMWLK4,JIMWLK5,JIMWLK6,JIMWLK7,JIMWLK8,AKLL,KOLU11,KOLUD,BA05,SMITH,KLW,KLLL1,KLLL2,kl,LEPR,LE1,LE2}, the problem of summation of the Pomeron loops has  not been solved.  

\begin{figure}[ht]
    \centering
  \leavevmode
      \includegraphics[width=8cm]{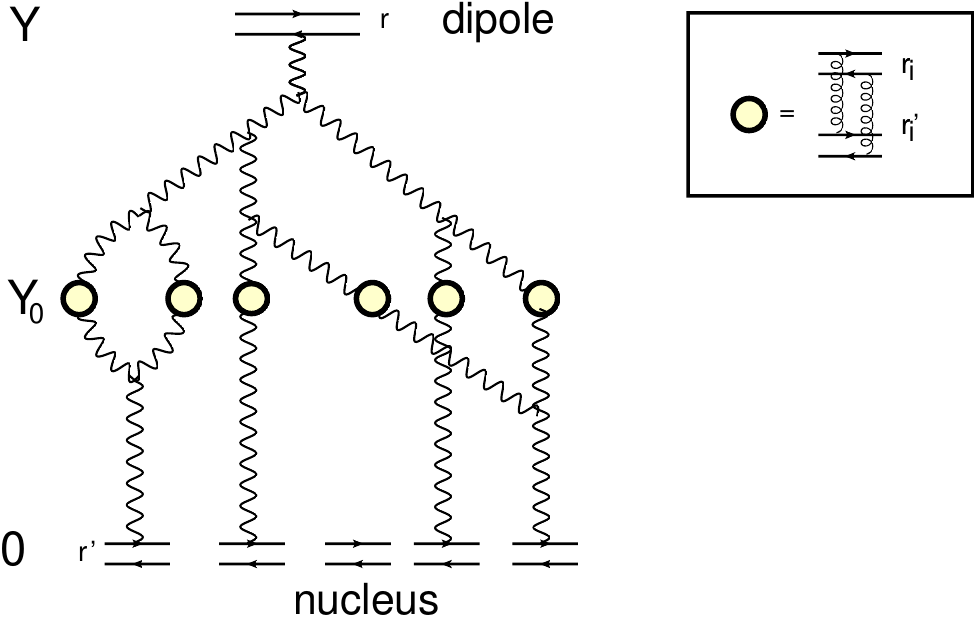}  
      \caption{Summing  large Pomeron loops for dipole-nucleus scattering. The wavy lines denote the  BFKL Pomeron exchanges.  The circles denote the amplitude $\gamma$ in the Born approximation of perturbative QCD. A nucleus is viewed as the bag of dipoles with the size $r'$. }
\label{mpsi}

   \end{figure}

In our recent paper\cite{LEDIDI} we  have suggested the way to sum the large Pomeron loops for dipole-dipole scattering deeply in the saturation region. 
In this paper  the large Pomeron loops is summed  using the $t$-channel unitarity, which has been rewritten in the convenient form for the dipole approach to CGC in Refs.\cite{MUSA,Salam,IAMU,IAMU1,KOLEB,MUDI,LELU,KO1,LE11}(see \fig{mpsi}).
 The analytic expression  for arbitrary $Y_0$ takes the form
       \cite{LELU,KO1,LE1}:  \bea \label{MPSI}
     && A\Lb Y, r, r' ;  \vec{b}\Rb\,=\\
     &&\,\sum^\infty_{n=1}\,\Lb -1\Rb^{n+1}\,n!\int  \prod^n_i d^2 r_i\,d^2\,r'_i\,d^2 b'_i 
     \int \!\!d^2 \delta b_i\, \gamma^{BA}\Lb r_i,r'_i, \vec{b}_i -  \vec{b'_i}\equiv \delta \vec{b} _i\Rb 
    \,\,\rho^P_n\Lb Y - Y_0,r,b,  \{ \vec{r}_i,\vec{b}_i\}\Rb\,\rho^T_n\Lb Y_0,r', b', \{ \vec{r}'_i,\vec{b}'_i\}\Rb \nn
      \eea
  $\gamma^{BA}$ is the scattering amplitude of two dipoles in the Born approximation of perturbative QCD.  The dipole  densities: $\rho^P_i \Lb Y -Y_0 , r, b, \{ \vec{r}_i,\vec{b}_i\}\Rb$  for projectile and $\rho^T_i\Lb Y_0 , r', b', \{ \vec{r}'_i,\vec{b}_i\}\Rb$  for target, have been introduced in Ref.\cite{LELU}  as follows:
\beq \label{PD}
\rho_n( Y\,-\,Y_0, r, b, r_1, b_1\,
\ldots\,,r_n, b_n)\,=\,\frac{1}{n!}\,\prod^n_{i =1}
\,\frac{\delta}{\delta
u_i } \,Z\left(Y\,-\,Y_0;\,[u] \right)|_{u=1}
\eeq
  where  the generating functional $Z$ is
  \beq \label{Z}
Z\Lb Y, \vec{r},\vec{b}; [u_i]\Rb\,\,=\,\,\sum^{\infty}_{n=1}\int P_n\Lb Y,\vec{r},\vec{b};\{\vec{r}_i\,\vec{b}_i\}\Rb \prod^{n}_{i=1} u\Lb \vec{r}_i\,\vec{b}_i\Rb\,d^2 r_i\,d^2 b_i
\eeq
 where $u\Lb \vec{r}_i\,\vec{b}_i\Rb \equiv\,u_i$ is an arbitrary function and $P_n$ is the probability to have $n$ dipoles with the  given kinematics.
 The initial and  boundary conditions for the BFKL cascade  stem from one dipole has 
the following form for the functional $Z$:
\begin{subequations}
\bea
Z\Lb Y=0, \vec{r},\vec{b}; [u_i]\Rb &\,\,=\,\,&u\Lb \vec{r},\vec{b}\Rb;~~~~~~~~Z\Lb Y, r,[u_i=1]\Rb = 1;\label{ZIC}\\
\rho_1\Lb Y=0,   r,b, r_1,b_1\Rb\,\,&=&\,\,\delta^{(2)}\Lb \vec{r} - \vec{r}_1\Rb \delta^{(2)}\Lb \vec{b} - \vec{b}_1\Rb ;~~~~~\rho_n\Lb 
 Y=0, \vec{r},\vec{b}; [r_i, b_i]\Rb \,=\,0 ~ \mbox{at}~\,n\geq 2;\label{ZSR}
\eea
\end{subequations}
 In \eq{MPSI} $\vec{b}_i\,\,=\,\,\vec{b} \,-\,\vec{b'}_i$. 
 Comparing \eq{PD} and \eq{Z} one can see that  $\rho_n = M_n/n!$ where $M_n$ is the factorial moment for the scattering amplitude. Factorial moments are the natural   observables for the BFKL Pomeron calculus (see Ref.\cite{MUDI}).
 
The dipole densities for the projectile (fast dipole) have been found in Ref.\cite{LEDIDI}. For completeness of presentation we will discuss them in the next section. However we wish to stress here that they satisfy both the
 evolution equations  and  the recurrence relations derived in QCD (see Refs.\cite{LELU,LELU1,LE1}; and the analytic
 solution to the nonlinear BK equation of Ref.\cite{LETU}. This solution takes the form:
 \beq\label{I1}
 N^{\rm DIS} \Lb \tau= e^z= r^2 Q^2_s\Lb Y, b\Rb\Rb\,\,=\,\,1\,\,-\,\,C(z)\exp\Lb - \frac{z^2}{2\,\kappa}\Rb
 \eeq 
 where $z = \ln\Lb r^2 Q^2_s(Y)\Rb$\footnote{We will discuss the definition of $z$ as well as  the saturation momentum $Q_s\Lb Y\Rb$ in a bit more details  below in the next section.}. One can see that  this solution
  shows the geometric scaling behaviour \cite{GS}  being a function of one variable. Function $C\Lb z\Rb$ is a smooth function  which can be considered as a constant in our approach.

   As one can see from \eq{I1}   it turns out that 
 BK equation  leads to a new dimensional scale: saturation momentum\cite{GLR}  which has the following $Y$ dependence\cite{GLR,MUT,MUPE}:
 \beq \label{QS}
 Q^2_s\Lb Y, b\Rb\,\,=\,\,Q^2_s\Lb Y=0, b\Rb \,e^{\bas\,\kappa \,Y -\,\,\frac{3}{2\,\gamma_{cr}} \ln Y }
 \eeq 
 where $Y=0$ is the initial value of rapidity and $\kappa$ and $\gamma_{cr}$   are determined by the following equations:
  \beq \label{GACR}
\kappa \,\,\equiv\,\, \frac{\chi\Lb \gamma_{cr}\Rb}{1 - \gamma_{cr}}\,\,=\,\, - \frac{d \chi\Lb \gamma_{cr}\Rb}{d \gamma_{cr}}~
\eeq
where $\chi\Lb \gamma\Rb$ is given by
\beq \label{CHI}
\omega\Lb \bas, \gamma\Rb\,\,=\,\,\bas\,\chi\Lb \gamma \Rb\,\,\,=\,\,\,\bas \Lb 2 \psi\Lb 1\Rb \,-\,\psi\Lb \gamma\Rb\,-\,\psi\Lb 1 - \gamma\Rb\Rb\eeq 
$\psi(z)$ is the Euler $\psi$ function (see Ref.\cite{RY} formulae {\bf 8.36}).

In the next section we show how to reconcile this solution with the fact that the BK equation is summing the 'fan' diagrams of \fig{gen}-a in the BFKL Pomeron calculus. We present \eq{I1} as a sum of many Pomerons exchanges and in doing so we find the dipole densities $\rho_n$.

  In section III we estimate the scattering amplitude using \eq{MPSI}  and the dipole densities that we found in section II. 
 In conclusion we summarize our results and discuss possible flaw in our approach,

    \begin{boldmath}
    \section{Dipole densities $\rho_n\Lb r,b, \{r_i,b_i\}\Rb$ }

    \end{boldmath}
    
    
      \begin{boldmath}    
        \subsection{ The BFKL Pomeron Green's function( $\rho_1\Lb r,b, r_1b_1\Rb$) in the saturation region }     
    \end{boldmath}
    
The Green's function for the BFKL Pomeron exchange satisfy the linear equation\cite{BFKL} which has the  form:

 \beq \label{GF1}
  \frac{\partial\,G_{\pom}\Lb Y,  \vec{r}; \vec{b} \Rb }{\partial\,Y} = \bas \intl \frac{d^2 r'}{2\,\pi} \frac{r^2}{r'^2 \Lb \vec{r} -\vec{r}'\Rb^2}\,\Bigg\{ G_{\pom}\Lb Y,   \vec{r}'; \vec{b} - \h\Lb \vec{r} - \vec{r}'\Rb \Rb+G_{\pom}\Lb Y,  \vec{r}  -  \vec{r}'; \vec{b} - \h\vec{r}' \Rb -G_{\pom}\Lb Y,  \vec{r}; \vec{b} \Rb\Bigg\} \eeq
   Solution to this equation is the sum over the eigenfunction $ \phi_\gamma\Lb \vec{r} , \vec{r}_1, \vec{b}\Rb $ with the eigenvalues $\bas \chi\Lb \gamma\Rb$:
    \beq \label{GF2}   
   G_{\pom}\Lb Y,  \vec{r}, \vec{r}_1; \vec{b} \Rb  \,\,=\,\, \!\!\!\intl^{\epsilon + i \infty}_{\epsilon - i \infty} \!\!\!\frac{d \gamma}{2\,\pi\,i} e^{ \bas \chi\Lb \gamma\Rb\,Y}\, \phi_\gamma\Lb \vec{r} , \vec{r}_1, \vec{b}\Rb  \phi_{in}\Lb r_1\Rb
   \eeq
   The eigenfunction  $ \phi_\gamma\Lb \vec{r} , \vec{r}_1, \vec{b}\Rb $ (the scattering amplitude of two dipoles with sizes $r$ and $r_1$) is equal to \cite{LIP}
   \beq \label{XI}
\phi_\gamma\Lb \vec{r} , \vec{r}_1, \vec{b}\Rb\,\,\,=\,\,\,\Lb \frac{
 r^2\,r_1^2}{\Lb \vec{b}  + \h(\vec{r} - \vec{r}_1)\Rb^2\,\Lb \vec{b} 
 -  \h(\vec{r} - \vec{r}_1)\Rb^2}\Rb^\gamma =e^{\gamma\,\xi_{r,r_1} }~~\mbox{with}\,\,0 \,<\,Re\gamma\,<\,1
 \eeq

$\phi_{in}\Lb r_1\Rb $ can be found from the initial conditions: at $Y=0$ $G_{\pom}$ has to describe the scattering amplitude in the Born approximation of perturbative QCD ( the  exchange of two gluons between dipoles with sizes $r$ and $r_1$ , see insertion in \fig{mpsi}). The eigenvalue  $ \bas \chi\Lb \gamma\Rb$ is given by \eq{CHI}.

At high energies (deeply in the saturation region) the exchange of one BFKL Pomeron can written in the following way\cite{GLR,MUT}:
\beq \label{DD1}
G_{\pom}\Lb z \Rb  = N_0 e^{\bg \,z}\eeq
where $\bg = 1 - \gamma_{cr}$ , $N_0$ is a constant,  $z$ in \eq{DD1} is defined as 
\beq \label{zz}
z\,\,=\,\,\bas \frac{\chi\Lb \bg\Rb}{\bg} \Lb Y \,-\,Y_0\Rb \,\,+\,\,\xi_{r,r_1}
\eeq 
where
$\xi_{r,r_1}$ is given by \eq{XI}.  
   As one can see from \eq{DD1}   it turns out that 
 BK equation  leads to a new dimensional scale: saturation momentum\cite{GLR}  which  $Y$ dependence is given by \eq{QS}\cite{GLR,MUT,MUPE}.

The Pomeron Green's function satisfies the $t$-channel unitarity\cite{BFKL}  
 analytically  continued to the $s$-channel, and can be re-written as the integration over two reggeized gluons at fixed momentum $Q_T$  that carries each Pomeron\cite{GLR,MUDI}(see \fig{gen}-c) : 

   \beq \label{TU1}  
G_{\pom}\Lb Y, Q_T,  r, r'\Rb \,\,=\,\,\int\frac{ d^2 k_T}{(2\,\pi)^2}\, G_{\pom}\Lb Y - y',Q_T,  r, k_T \Rb   G_{\pom}\Lb y', Q_T,  r' ,  k_T\Rb 
\eeq
where 

\beq \label{GFMR}
 G_{\pom}\Lb Y - y', Q_T,  r, r_1\Rb\,\,\,=\,\, r'^2 \int \frac{d^2 k_T}{(2\,\pi)^2} e^{i \vec{k}_T \cdot \vec{r}_1}    G_{\pom}\Lb Y - y',Q_T,  r, k_T\Rb
 \eeq     
  \eq{TU1} can be re-written through $G_{\pom}\Lb Y - y', r, r', Q_T\Rb$ in the form of \eq{MPSI}(see \fig{gen}-c):
  \begin{subequations}
   \bea \label{TU2}
G_{\pom} \Lb Y, r,r' ;  \vec{b}\Rb \,\,&=&\,\,\intl d^2 r_1 d^2 b_1 d^2r'_1 d^2 b'_1
\rho_1\Lb Y - Y_0; r, b -b_1, r_1\Rb \gamma^{BA}\Lb r_1 ,r'_1,\delta b= \vec{b}_1 - \vec{b}'_1\Rb \rho_1\Lb  Y_0; r',0,b'_1, r'_1\Rb\label{TU2}\\
&\,\,=\,\,&
 \,\,\frac{1}{4\,\pi^2}\int \frac{d^2 r_1}{r^4_1} \,d^2 b' 
   \,   G_{\pom} \Lb Y - Y_0, r, r_1;  \vec{b}\ - \vec{b}' \Rb \,\, G_{\pom} \Lb Y_0, r',  r_1;  \vec{b}' \Rb \label{TU20} 
   \eea 
    \end{subequations}

~
   The first equation is derived in Ref.\cite{MUDI} (see also Refs.\cite{MUSA,Salam}) while  the second in Ref.\cite{CLMSOFT}.

   ~~
   
   ~
   
   ~

    \begin{boldmath}

      \subsection{ $\rho^P_n\Lb r,b, \{r_i,b_i\}\Rb$ for projectile (fast dipole) }     
    \end{boldmath}
    

In Ref.\cite{LEDIDI} we found the dipole densities for the projectile (fast dipole with size $r$). They are equal to
\beq \label{DD5}
\rho^P_n\Lb Y - Y_0;  r,b, \{ r_i,b_i\}\Rb = C \,\intl^{\epsilon + i \infty}_{\epsilon - i \infty} \frac{d \omega}{2\,\pi\,i} e^{ \frac{ 
\bg\,\kappa}{2}\,\omega^2}\frac{1}{n!} \frac{\Gamma\Lb \omega +n\Rb}{\Gamma\Lb \omega \Rb}  \prod^n_{i=1} \frac{G_{\pom}\Lb z_i\Rb}{N_0}
\eeq
where
\beq \label{zi}
z_i \,\,=\,\,\,\,\bas \frac{\chi\Lb \bg\Rb}{\bg} \Lb Y \,-\,Y_0\Rb \,\,+\,\,\xi_{r,r_i}
\eeq
 with $\xi_{r, r_i}$ from \eq{XI} in which $r_1 $ is replaced by   $r_i$.
 
 \eq{DD5}, as it is shown in Ref.\cite{LEDIDI}, 
   stems from our attempts to reconcile the exact solution to the Balitsky-Kovchegov (BK) equation, which describes the rare fluctuation in the dipole-target scattering, with the fact that this equation sums the 'fan' Pomeron diagrams.  For completeness of presentation we will show below how \eq{DD5} reproduces the solution of Ref.\cite{LETU} for the BK equation.

Using these densities  the scattering matrix 
 $S\Lb z\Rb$ can be written as \cite{KOLEB,K,LELU}
\beq \label{DD4}
S\Lb z\Rb\,\,=\sum^{\infty}_{n=0} (-1)^n \int \prod^n_{i=1} \gamma\Lb r_i, r'; b_i\Rb\,d^2r_i\,d^2 b_i\,\,\rho_n\Lb Y, r,b, \{ r_i,b_i\}\Rb 
\eeq
where $\gamma$ is the amplitude of interaction of the dipole with rapidity $Y=0$ and the size $r_i$ with the target  of size $r'$ (see \fig{gen}-a and \fig{mpsi}) in the Born approximation of perturbative QCD.    $z = z_i $ for $r_i = r'$.  In derivation of \eq{DD4} we use that $S\Lb Y, r, r',b\Rb= Z\Lb Y, r,b,[1 -\gamma\Lb r_i, r'; b_i\Rb]\Rb$  and the definition of $\rho_n$ in \eq{PD} (see Ref.\cite{LELU} for details). Generally speaking $S$ is a function of $Y, r, r' ,b$. The fact that $S$ is a function of only one variable  stems from \eq{DD5} for the dipole densities or,  more generally,  from the geometric scaling behaviour of the scattering amplitude \cite{GS}.

It is instructive to note that \eq{DD4} follows directly from \eq{MPSI} if we choose $Y_0 =0$ in \eq{MPSI} and  consider that
\beq \label{DD41}
\rho^T_n\Lb Y_0=0, r', b',  \{ \vec{r}'_i,\vec{b}'\}\Rb \,\,=\,\,\frac{S^n_A\Lb b'\Rb}{n!}\prod^n_{i=1} \delta^{(2)}\Lb \vec{r}' - \vec{r}_i\Rb \delta^{(2)}\Lb \vec{b}' - \vec{b}_i\Rb ;
\eeq
 \eq{DD41} has simple meaning (see \eq{ZSR}), that  we consider a nucleus as a bag of dipoles with size $r'$ which do not interact with each other. Factor $S^n_A\Lb b' \Rb$ describes the probability to find $n$- nucleons (dipoles) in a nucleus. Function $S\Lb b'\Rb$ we will discuss below.

\eq{DD4} can be easily derived from the McLerran-Venugopalan formula for the scattering amplitude \cite{MV} at initial rapidity $Y_0=0$  in which
\beq \label{DD42}
\gamma\Lb r_i,r'; b\Rb\,\,=\,\, S_A\Lb b \Rb \int\!\! d \,\delta b\,\gamma^{BA}\Lb  r_i,r'; \delta b_ i\Rb
\eeq
where $S_A\Lb b \Rb$ is the nucleus profile function which is equal to
\beq \label{DD43}
S_A\Lb b \Rb\,\,=\,\,\int^{\infty}_{-\infty}\!\!\!\! d z \rho_A\Lb \sqrt{b^2+z^2}\Rb 
\eeq
with $\rho_A$ is the density of nucleons in the nucleus A.

Plugging \eq{DD5} into \eq{DD4}  and integrating over $r_i,b_i$, using \eq{TU2},  we obtain the following expression for the scattering matrix (S-matrix)\footnote{We use the standard notation for the scattering matrix ($S(z)$ and  mean S-matrix S(z) or nucleus thickness function  ($S_A\Lb b\Rb$). Hopefully this will not confuse a reader.}:
\begin{subequations}
\bea
S\Lb z\Rb\,\,&=& \,\, C \,\intl^{\epsilon + i \infty}_{\epsilon - i \infty} \frac{d \omega}{2\,\pi\,i} e^{ \frac{ 
\bg^2\,\kappa}{2}\,\omega^2}   \sum^\infty_{n = 0} \frac{(-1)^n}{n!} \frac{\Gamma\Lb \omega +n\Rb}{\Gamma\Lb \omega \Rb}\Lb \frac{G_{\pom}\Lb z\Rb}{N_0}\Rb^n \,\,
= \,\,C\intl^{\epsilon + i \infty}_{\epsilon - i \infty} \frac{d \omega}{2\,\pi\,i} e^{ \frac{ 
\bg^2\,\kappa}{2} \omega^2} \Lb 1 + \frac{1}{N_0} G_{\pom}\Lb z \Rb\Rb^{-\omega} \label{DD3}\\
&\,\,=&\,\,\frac{C}{\sqrt{2\,\pi\,\bg^2\,\kappa}}\exp\Lb-\,\frac{1}{2\,\bg^2\,\kappa}\ln^2\Lb 1 + \frac{G_{\pom}\Lb z\Rb}{N_0} \Rb\Rb\,\,
\xrightarrow{z \gg 1} \frac{C}{\sqrt{2\,\pi\,\bg^2\,\kappa}}e^{ - \frac{z^2}{ 2\, \kappa}} \,\,\label{DD2} \eea
\end{subequations}
Note that integrating the BFKL Pomeron contributions over $r_i$ and $b_i$  using \eq{TU20}, we obtain the function of the same argument for each term in the sum: $z = z_i$ of \eq{zi} for  $r_i = r'$.

One can see that \eq{DD3} gives you the scattering amplitude that satisfies \eq{I1} as a  sum of 'fan' diagrams of the BFKL Pomeron calculus\cite{KOLEB,GLR,BRN,BRAUN} (see \fig{gen}-a). Each term of this sum  is the contribution of the exchange of $n$ Pomerons
$G_{n \pom} $ ( $ G_{n \pom} = \Lb G_{\pom}\Lb z \Rb \Rb^n$). 

However, in the case the  interaction with a nucleus we have to return to \eq{DD1}. 
 Indeed, from \eq{DD42} one can see that the exchange of one BFKL Pomeron can be written as
\beq \label{DD6}
G^A_{\pom} \Lb Y, r,  r', b\Rb\,\,=\,\,S_A\Lb b\Rb G_{\pom} \Lb Y, r, r', b\Rb\,\,=\,\,\,S_A\Lb b\Rb G_{\pom}\Lb z \Rb \,\, =\,\,S_A\Lb b\Rb\,N_0 e^{\bg \,z}=
N_0 e^{\bg \,z_A}
\eeq
with
\beq \label{DD7}
z_A\,\,=\,\,\bas \frac{\chi\Lb \bg\Rb}{\bg} \Lb Y \,-\,Y_0\Rb \,\,+\,\,\xi_{r,r'}\,\,+\,\frac{1}{\bg} \ln\Lb S_A\Lb b\Rb\Rb
\eeq

    ~

~

    \begin{boldmath}

      \subsection{ $\rho^T_n\Lb r,b, \{r_i,b_i\}\Rb$ for target (nucleus) }     
    \end{boldmath}
    

  Looking in \fig{mpsi} one can see that dipole  cascades with rapidities less that $Y_0$ include only annihilation of two dipoles to one ( $2 \pom \to \pom$ vertices).  Hence each nucleon of the target develops  its own  dipole  cascade, making $ \rho^T_n\Lb r,b, \{r_i,b_i\}\Rb$    is equal to
  \bea \label{DA1}
 &&  \rho^T_n\Lb Y_0,  r,b, \{r_i,b_i\}\Rb   \,\,=\,\, \sum^{n}_{j=1}\!\!\!\! \!\!\!\!\!\!\!\!\underbrace{\frac{1}{j!} S^j_A\Lb b\Rb}_{probability \,to\,find\,\,j-nucleons}\!\!\!\! \underbrace{\sum^{k_1=n - j +1,\dots \,,k_j=n - j +1}_{k_1=1,\dots\,,k_j=1}\,\delta_{\sum _{i=1}^j k_i = n}}_{summimg\,all\,k_i}\nn\\
&&
\rho_{k_1} \Lb Y_0;  r,' b', r_1, \dots\,, r_{k_1} \Rb   \times \dots \times \underbrace{\rho_{k_i} \Lb Y_0;  r,' b', r_{k_{i-1}+1} , \dots\,r_{k_i} \Rb}_{dipole\,density\, produced\, by \,i-th\,nucleon} \times \dots \times \rho_{k_j} \Lb Y_0;  r,' b', r_{k_{j-1}+1} , \dots\,r_{k_j} \Rb 
\eea  
This formula describes the simple fact that each i-th  nucleon produces $k_i$ dipoles and $\sum_{i=1}^j k_i=n$, where   $n$ is the total number of dipoles at $Y_0$. Each $k_i$ runs from $k_i=1$ to $k_i = n - j +1$ with additional restriction that $\sum_{i=1}^j k_i=n$ as it is written in the sum of the second factor of \eq{DA1}.

In this equation we took into account \eq{DD4} and \eq{DD41} and that all $b' \to b$ for large nuclei,  since all Pomeron interactions occur at $| \vec{b} - \vec{b}_i | \sim r_i \ll b$.   In particular, we can integrate over $\delta b_i $ in \eq{MPSI} replacing $\gamma^{BA}$ by $\sigma^{BA} = \int d^2 \delta b_i\, \gamma^{BA}\Lb r_i,r'_i, \vec{b}_i -  \vec{b'_i}\equiv \delta \vec{b} _i\Rb $.
     Using \eq{DD5} we can obtain that
 \beq \label{DA2}
  \rho^T_n\Lb r,b, \{r_i,b_i\}\Rb   \,\,=\,\,\prod^n_{i=1} \frac{G_{\pom}\Lb z_i\Rb}{N_0}
 \sum^{n}_{j=1}  \frac{1}{j!} S^j_A\Lb b\Rb\!\!\!\!\!\!\!\!\!\!\!\!\!\!\!\sum^{k_1=n - j+1,\dots\,,k_j=n-j + 1}_{k_1=1,\dots\,,k_j=1}\!\!\!\!\!\!\!\!\!\!\!\!\!\!\!\!\!\!\!\!\delta_{\sum _{i=1}^j k_i = n}  \prod^{j}_{i=1} 
C \,\intl^{\epsilon + i \infty}_{\epsilon - i \infty} \frac{d \omega_i}{2\,\pi\,i} e^{ \frac{ 
\bg\,\kappa}{2}\,\omega_i^2}\frac{1}{k_i!} \frac{\Gamma\Lb \omega_i +k_i\Rb}{\Gamma\Lb \omega_i \Rb} \eeq

In this paper we use \eq{DA2} for  $\rho^T_n\Lb r,b, \{r_i,b_i\}\Rb$ replacing
$\Gamma\Lb \omega_i +k_i\Rb$ by the integrals over $t_i$, viz.:
 \beq \label{DA21}
 \Gamma\Lb \omega_i +k_i\Rb\,\,=\,\,\intl^\infty_0 d t_i t_i^{ \omega_i +k_i -1} e^{- t_i}
 \eeq  
 In doing so we obtain
 
  \bea \label{DA22}
  &&\rho^T_n\Lb r,b, \{r_i,b_i\}\Rb   \,\,=\\
   &&\,\,\prod^n_{i=1} \frac{G_{\pom}\Lb z_i\Rb}{N_0}
 \sum^{n}_{j=1}  \frac{1}{j!} S^j_A\Lb b\Rb\!\!\!\!\!\!\!\!\!\!\!\!\!\!\!\sum^{k_1=n - j+1,\dots\,, k_j=n-j+1}_{k_1=1,\dots\,, k_j=1}\!\!\!\!\!\!\!\!\!\!\!\!\!\!\!\!\!\!\!\!\delta_{\sum _{i=1}^j k_i = n}  \prod^{j}_{i=1} 
C \,\intl^{\epsilon + i \infty}_{\epsilon - i \infty} \frac{d \omega_i}{2\,\pi\,i} e^{ \frac{ 
\bg\,\kappa}{2}\,\omega_i^2}\frac{1}{k_i!} \frac{1}{\Gamma\Lb \omega_i \Rb} 
\intl^\infty_0 d t_i t^{ \omega_i +k_i -1} e^{- t_i}\nn
\eea

In the appendix we discuss other representation of these densities which could be useful.
    ~
    
    ~
      
     \begin{boldmath}
     \section{Scattering amplitude}

      \end{boldmath}

\begin{figure}[ht]
    \centering
  \leavevmode
  \begin{tabular}{c c} 
      \includegraphics[width=10cm]{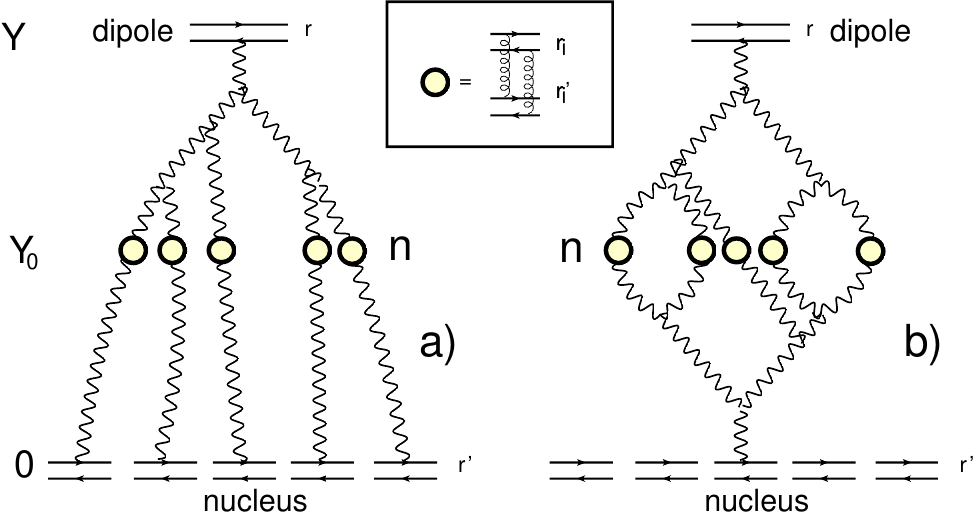} &  \includegraphics[width=5cm]{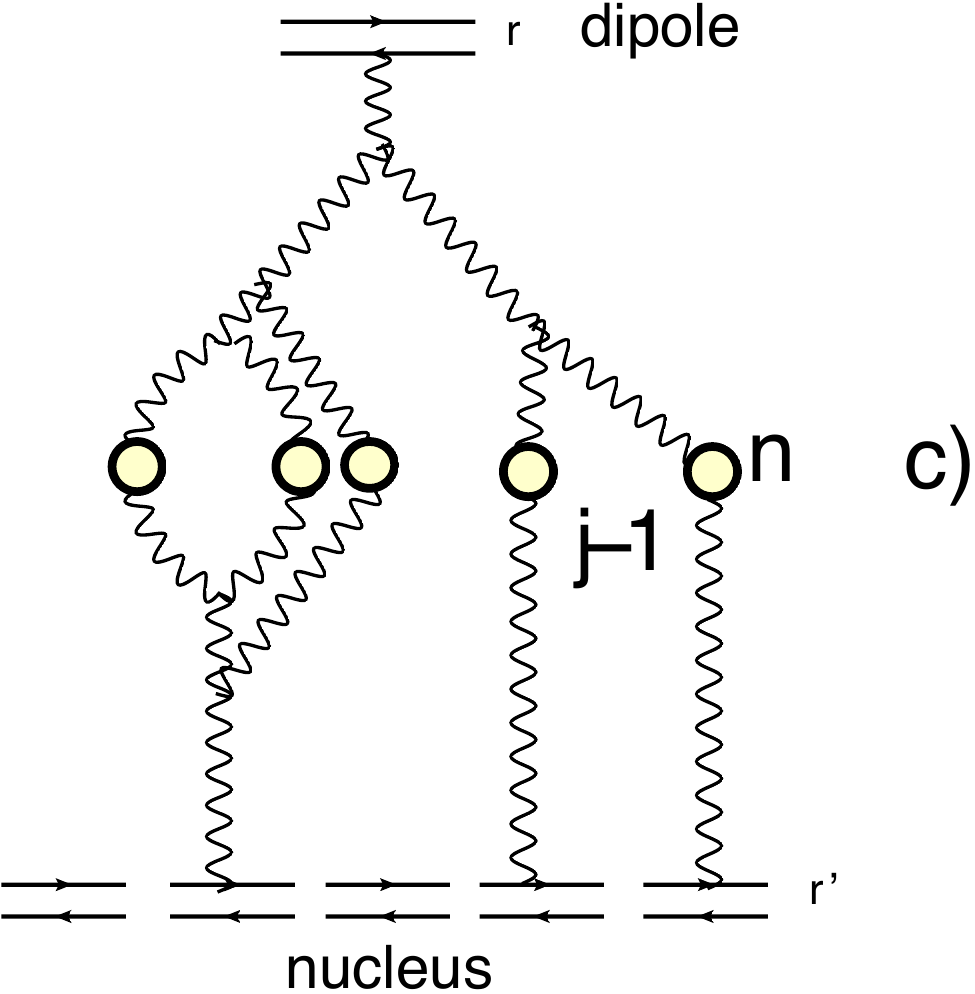} \\
      \end{tabular}
           \caption{Summing  large Pomeron loops for dipole-nucleus scattering. 
      \fig{2exmpl}-a: the  dipole cascade of the fast dipole interacts with one dipole from the each dipole  cascade  generated by the nucleons  (j =n in \eq{DA1}).   \fig{2exmpl}-b: the  dipole  cascade of the fast dipole interacts with the  dipole  cascade  generated by one  nucleon in a nucleus (j =1 in \eq{DA1}).
 \fig{2exmpl}-c: The main contribution for arbitrary values of $j$.     
 The wavy lines denote the  BFKL Pomeron exchanges.  The circles denote the amplitude $\gamma$ in the Born approximation of perturbative QCD. A nucleus is viewed as the bag of dipoles with the size $r'$. }
\label{2exmpl}  
   \end{figure}

      
      We start to discuss the scattering amplitude  considering two limited cases, shown in \fig{2exmpl}.
      
      ~

     \begin{boldmath}
     \subsection{j = n}

      \end{boldmath}


      ~

 In  \fig{2exmpl}-a all dipoles with rapidity $Y_0$ comes from different nucleons. In this case

      \beq \label{SA1}
 \rho^T_n\Lb Y_0,  r',b', \{r'_i,b'_i\}\Rb \,\,=\,\,\frac{1}{n!} S^n_A\Lb b' \Rb\prod_{i=1}^n \frac{G\Lb z'_i\Rb}{N_0} 
 \eeq
 
 Plugging this equation together with \eq{DD5} for the dipole densities of the projectile in \eq{MPSI} and using \eq{TU2} one can see  that for the scattering amplitude we obtain \eq{DD3}. Hence for this case we have the same solution as for the BK nonlinear equation.

 ~
 
 ~

     \begin{boldmath}
     \subsection{j = 1}

      \end{boldmath}

  \fig{2exmpl}-b shows the case when the fast dipole interacts only with one nucleon in the nucleus. The scattering amplitude is equal to
  \beq \label{SA2}
 N\Lb d A\Rb\,\,=\,\,S_A\Lb b \Rb N\Lb Y,  r, r'  \Rb    
 \eeq
 where $N$ is dipole-dipole scattering amplitude that has been discussed in Ref.\cite{LEDIDI}.
 For completeness presentation as well as to  illustrate   the methods that we use here , we    re-derive \eq{SA2}.

Using \eq{DD5} for the dipole densities we can find the sum of large Pomeron loops using  the $t$-channel unitarity constraints of \eq{MPSI}. It takes the form:
\bea \label{SA3}
&&S\Lb z\Rb =S_A\Lb b \Rb C^2\\
&&\sum^\infty_{n=0} \frac{\Lb - 1\Rb^n}{n!}\!\!\!  \intl^{\epsilon + i \infty}_{\epsilon - i \infty}\!\!\! \frac{ d \omega}{2\,\pi\,i} \!\!\!\intl^{\epsilon + i \infty}_{\epsilon _ i \infty}\!\!\!  \frac{d \omega_1}{2\,\pi\,i} e^{ \frac{ \bg^2 \kappa}{2}\Lb \omega^2 + \omega^2_1\Rb}\frac{ \Gamma\Lb \omega+n\Rb}{\Gamma\Lb \omega\Rb}\frac{ \Gamma\Lb \omega_1+n\Rb}{\Gamma\Lb \omega_1\Rb}\intl\!\!\! d^2 r_i d^2 r'_i d^2b_i d^2b'_i\prod^n_{i=1} 
\frac{G_{\pom}\Lb z_i\Rb}{N_0}\gamma^{BA}\Lb r_i,r'_i,\delta b\Rb\frac{G_{\pom}\Lb z'_i\Rb}{N_0}\nn\eea

In   \eq{SA3}  $z_i$ is given in \eq{zi} while $z'_i \,\,=\,\,\,\,\bas \frac{\chi\Lb \bg\Rb}{\bg} \Lb Y_0\Rb \,\,+\,\,\xi_{r',r'_i}$. $\xi_{r',r'_i}$ is equal to $\xi_{r,r_i}$ of \eq{XI} where $r$  and $r_i$ are  replaced by $r'$ and $r'_i$, respectively.

  Using \eq{TU2} we reduce \eq{SA3} to the following expression:
  \beq \label{SA31}
S\Lb z'\Rb =S_A\Lb b \Rb
C^2\sum^\infty_{n=0} \frac{\Lb - 1\Rb^n}{n!}\!\!\!  \intl^{\epsilon + i \infty}_{\epsilon - i \infty}\!\!\! \frac{ d \omega}{2\,\pi\,i} \!\!\!\intl^{\epsilon + i \infty}_{\epsilon _ i \infty}\!\!\!  \frac{d \omega_1}{2\,\pi\,i} e^{ \frac{ \bg^2 \kappa}{2}\Lb \omega^2 + \omega^2_1\Rb}\frac{ \Gamma\Lb \omega+n\Rb}{\Gamma\Lb \omega\Rb}\frac{ \Gamma\Lb \omega_1+n\Rb}{\Gamma\Lb \omega_1\Rb}\Lb \frac{G_{\pom}\Lb z'\Rb}{N_0}\Rb^n
\eeq
 with 
 \beq \label{SA4}
 z' \,\,= \,\,\,\,\bas \frac{\chi\Lb \bg\Rb}{\bg} \,Y \,\,+\,\,\xi_{r,r'}    
 \eeq   
  $\xi_{r,r'}$ is equal to $\xi_{r,r_1}$ of \eq{XI} where $r$  and $r_1$ are  replaced by $r$ and $r'$, respectively.
  
  Plugging    \eq{DA21}  into  \eq{SA4}  we obtain:
    \beq \label{SA5}
S\Lb z'\Rb =C^2\,S_A\Lb b \Rb
\sum^\infty_{n=0} \frac{\Lb - 1\Rb^n}{n!}\!\!\!  \intl^{\epsilon + i \infty}_{\epsilon - i \infty}\!\!\! \frac{ d \omega}{2\,\pi\,i} \!\!\!\intl^{\epsilon + i \infty}_{\epsilon -i \infty}\!\!\!  \frac{d \omega_1}{2\,\pi\,i} e^{ \frac{ \bg^2 \kappa}{2}\Lb \omega^2 + \omega^2_1\Rb}\intl^\infty_0 d t\,  e^{-t}  t^{\omega_1 -1}  \frac{ \Gamma\Lb \omega+n\Rb}{\Gamma\Lb \omega\Rb\Gamma\Lb \omega_1\Rb}\Lb t\, \frac{G_{\pom}\Lb z'\Rb}{N_0}\Rb^n
\eeq
  After summation over $n$ we have
          \bea \label{SA6}
S\Lb z'\Rb &=&C^2 S_A\Lb b \Rb
 \intl^{\epsilon + i \infty}_{\epsilon - i \infty}\!\!\! \frac{ d \omega}{2\,\pi\,i} \!\!\!\intl^{\epsilon + i \infty}_{\epsilon - i \infty}\!\!\!  \frac{d \omega_1}{2\,\pi\,i} e^{ \frac{ \bg^2 \kappa}{2}\Lb \omega^2 + \omega^2_1\Rb}\frac{ 1}{\Gamma\Lb \omega_1\Rb}\intl^\infty_0 d t  e^{-t} \Lb 1\,+\,t  \frac{G_{\pom}\Lb z'\Rb}{N_0} \Rb^{- \omega }\,t^{\omega_1 -1}\nn\\
 &\xrightarrow{G_{\pom}\Lb z'\Rb\,\gg\,1}&C^2 S_A\Lb b \Rb
 \intl^{\epsilon + i \infty}_{\epsilon - i \infty}\!\!\! \frac{ d \omega}{2\,\pi\,i} \!\!\!\intl^{\epsilon + i \infty}_{\epsilon - i \infty}\!\!\!  \frac{d \omega_1}{2\,\pi\,i} e^{ \frac{ \bg^2 \kappa}{2}\Lb \omega^2 + \omega^2_1\Rb}\frac{ 1}{\Gamma\Lb \omega_1\Rb}\intl^\infty_0 d t  e^{-t} \exp\Lb -\omega\, \ln\Lb \frac{G_{\pom}\Lb z'\Rb}{N_0}\Rb\Rb\,t^{\omega_1 - \omega  -1} \nn\\
 &=&C^2 S_A\Lb b \Rb
 \intl^{\epsilon + i \infty}_{\epsilon - i \infty}\!\!\! \frac{ d \omega}{2\,\pi\,i} \!\!\!\intl^{\epsilon + i \infty}_{\epsilon - i \infty}\!\!\!  \frac{d \omega_1}{2\,\pi\,i} e^{ \frac{ \bg^2 \kappa}{2}\Lb \omega^2 + \omega^2_1\Rb}\frac{ \Gamma\Lb \omega_1 - \omega\Rb}{\Gamma\Lb \omega_1\Rb} \exp\Lb -\omega\, \ln\Lb \frac{G_{\pom}\Lb z'\Rb}{N_0}\Rb\Rb  
\eea
Closing contour of integration over  $\omega_1$ on the poles of $\Gamma\Lb \omega - \omega_1\Rb$  one can see that the main contribution gives the pole $\omega - \omega_1$ while all other poles lead to the amplitude that decreases as $\exp\Lb -\bg\,n \,z'\Rb$ for the pole $\omega_1= \omega - n$. Taking integral over $\omega$ we obtain:
 The resulting $S\Lb z'\Rb $ takes the form which has been found in Ref.\cite{LEDIDI}:
  \beq \label{SA7}
  S\Lb z'\Rb = S_A\Lb b\Rb   C'^2  \exp\Lb - \frac{ z'^2}{4 \kappa}\Rb
  \eeq

  Hence,  we can conclude that the sum of the large Pomeron loops leads to the  S-matrix in perfect agreement with the estimates of 'rare' fluctuation, given in Ref.\cite{IAMU}.
         
       ~

     ~      
     \begin{boldmath}
     \subsection{j = 2}
 \end{boldmath}   
   
         ~

         ~
         
         Using  \eq{DD2} for $\rho^T_n$ we can rewrite \eq{MPSI} in the form:
      \bea \label{SA8}
&&S\Lb z\Rb =\h S^2_A\Lb b \Rb C^2\sum^\infty_{n=0} \frac{\Lb - 1\Rb^n}{n!}\!\!\!  \intl^{\epsilon + i \infty}_{\epsilon - i \infty}\!\!\! \frac{ d \omega}{2\,\pi\,i} \!\!\!\intl^{\epsilon + i \infty}_{\epsilon - i \infty}\!\!\!  \frac{d \omega_1}{2\,\pi\,i} 
\!\!\intl^{\epsilon + i \infty}_{\epsilon - i \infty}\!\!\!  \frac{d \omega_2}{2\,\pi\,i}
e^{ \frac{ \bg^2 \kappa}{2}\Lb\omega^2 +  \omega_1^2 + \omega^2_2\Rb}\frac{ \Gamma\Lb \omega+n\Rb}{\Gamma\Lb \omega\Rb}\nn\\
&&\sum_{k=1}^{n-1} \frac{n!}{k! (n -k)!} \frac{ \Gamma\Lb \omega_1+k\Rb}{\Gamma\Lb \omega_1\Rb}\frac{ \Gamma\Lb \omega_2+n-k\Rb}{\Gamma\Lb \omega_2\Rb}
\intl\!\!\! d^2 r_i d^2 r'_i d^2b_i d^2b'_i\prod^n_{i=1} 
\frac{G_{\pom}\Lb z_i\Rb}{N_0}\gamma^{BA}\Lb r_i,r'_i,\delta b\Rb\frac{G_{\pom}\Lb z'_i\Rb}{N_0}
\eea
Using  \eq{TU2} and replacing $\Gamma$ functions by the integrals over $t_i$ (see \eq{DA21}) we obtain
     \bea \label{SA9}
S\Lb z\Rb &=&\h S^2_A\Lb b \Rb C \sum^\infty_{n=0}
\Lb \frac{G_{\pom}\Lb z'\Rb}{N_0}\Rb^n\frac{(-1)^n}{n!} \!\!\!  \intl^{\epsilon + i \infty}_{\epsilon - i \infty}\!\!\! \frac{ d \omega}{2\,\pi\,i} \!\!\!\intl^{\epsilon + i \infty}_{\epsilon - i \infty}\!\!\!  \frac{d \omega_1}{2\,\pi\,i} 
\!\!\intl^{\epsilon + i \infty}_{\epsilon - i \infty}\!\!\!  \frac{d \omega_2}{2\,\pi\,i}
e^{ \frac{ \bg^2 \kappa}{2}\Lb\omega^2 +  \omega^2_1 + \omega^2_2\Rb}
\frac{ \Gamma\Lb \omega+n\Rb}{\Gamma\Lb \omega\Rb \Gamma\Lb \omega_1\Rb\Gamma\Lb \omega_2\Rb}\nn\\
& &\int^{\infty}_0 d t_1 \int^{\infty}_0 d t_2 e^{-t_1-t_2}
\underbrace{\sum_{k=1}^{n-1} \frac{n!}{k! \,(n -k)!} t_1^{\omega_1 + k -1} t_2^{\omega_1 +n -  k -1}}_{ (t_1 + t_2)^n - t_1^n - t_2^n}  
 \eea        
    Summing over $n$ we obtain:
     \bea \label{SA10}
S\Lb z\Rb &=&\h S^2_A\Lb b \Rb C \sum^\infty_{n=0}
\!\!\!  \intl^{\epsilon + i \infty}_{\epsilon - i \infty}\!\!\! \frac{ d \omega}{2\,\pi\,i} \!\!\!\intl^{\epsilon + i \infty}_{\epsilon - i \infty}\!\!\!  \frac{d \omega_1}{2\,\pi\,i} 
\!\!\intl^{\epsilon + i \infty}_{\epsilon - i \infty}\!\!\!  \frac{d \omega_2}{2\,\pi\,i} \,e^{ \frac{ \bg^2 \kappa}{2}\Lb\omega^2 +  \omega_1^2 + \omega^2_2 \Rb}
\frac{1}{\Gamma\Lb \omega_1\Rb\Gamma\Lb \omega_2\Rb\Gamma\Lb \omega_3\Rb}\exp\Lb -\omega\, \ln\Lb \frac{G_{\pom}\Lb z'\Rb}{N_0}\Rb\Rb\nn\\
& &\intl^{\infty}_0\!\!\!d t_1 \!\!\intl^{\infty}_0 \!\!\!d t_2 \,\,e^{ - t_1 - t_2 } 
\,\Bigg\{t_1^{\omega_1 -1}\,t_2^{ \omega_2  -1} \Lb t_1 + t_2\Rb^{-\omega}        
    -   \,t_1^{\omega_1 - \omega -1}\,t_2^{ \omega  -1}  - \,t_2^{\omega_2 - \omega -1}\,t_1^{ \omega_2  -1} \Bigg\}     \eea

    Taking integrals over $t_1$ and $t_2$ we reduce \eq{SA10} to the form:
    \begin{subequations}
     \bea 
S\Lb z\Rb &=&\h S^2_A\Lb b \Rb C\!\!\!  \intl^{\epsilon + i \infty}_{\epsilon - i \infty}\!\!\! \frac{ d \omega}{2\,\pi\,i} \!\!\!\intl^{\epsilon + i \infty}_{\epsilon - i \infty}\!\!\!  \frac{d \omega_1}{2\,\pi\,i} 
\!\!\intl^{\epsilon + i \infty}_{\epsilon - i \infty}\!\!\!  \frac{d \omega_2}{2\,\pi\,i}
e^{ \frac{ \bg^2 \kappa}{2}\Lb\omega^2 +  \omega_1^2 + \omega^2_2\Rb}
\exp\Lb -\omega\, \ln\Lb \frac{G_{\pom}\Lb z'\Rb}{N_0}\Rb\Rb\,\nn\\
& &\Bigg\{ \underbrace{   \frac{\Gamma\Lb  \omega_1 + \omega_2 - \omega\Rb}{\Gamma\Lb \omega_1 + \omega_2\Rb}}_{\rm I}  \,-\,\underbrace{ \frac{\Gamma\Lb  \omega_1  - \omega\Rb}{\Gamma\Lb \omega_1 \Rb}}_{\rm II}\,-\,\underbrace{ \frac{\Gamma\Lb   \omega_2 - \omega\Rb}{\Gamma\Lb \omega_2\Rb}}_{\rm III} \Bigg\} \label{SA11a}  \\
&=&\h S^2_A\Lb b \Rb C\Bigg\{ \underbrace{  C'  \exp\Lb - \frac{ z'^2}{3 \kappa}\Rb}_{\rm I} \,\,-\,\,\underbrace{ C'  \exp\Lb - \frac{ z'^2}{4 \kappa}\Rb}_{\rm II} \,-\,\,\underbrace{ C'  \exp\Lb - \frac{ z'^2}{4 \kappa}\Rb}_{\rm III} \Bigg\} \label{SA11b}\\
&\xrightarrow{z' \gg 1} &- S^2_A\Lb b \Rb  C'  \exp\Lb - \frac{ z'^2}{4 \kappa}\Rb\label{SA11c} \eea    
 \end{subequations}

       ~

     \begin{boldmath}
     \subsection{j = 3}
 \end{boldmath}   
   
         ~

         ~
         
       From \eq{MPSI}  the scattering amplitude in this case is equal to:
      \bea \label{SA12}
&&S\Lb z'\Rb =\frac{1}{3!} S^3_A\Lb b \Rb C^3\sum^\infty_{n=0} \frac{\Lb - 1\Rb^n}{n!}\!\!\!  \intl^{\epsilon + i \infty}_{\epsilon - i \infty}\!\!\! \frac{ d \omega}{2\,\pi\,i}  e^{ \frac{ \bg^2 \kappa}{2} \omega^2 }\frac{ \Gamma\Lb \omega+n\Rb}{\Gamma\Lb \omega\Rb}    \prod^3_{i=1}\, 
\!\!\intl^{\epsilon + i \infty}_{\epsilon - i \infty}\!\!\!  \frac{d \omega_i}{2\,\pi\,i}
e^{ \frac{ \bg^2 \kappa}{2} \omega_i^2 }\,\sum_{k_i=1}^{n-2} \frac{n!}{k_1!  k_2!(n -k_1 -k_2)!} \nn\\
&&\frac{ \Gamma\Lb \omega_1+k_1\Rb}{\Gamma\Lb \omega_1\Rb}\frac{ \Gamma\Lb \omega_2+k_2\Rb}{\Gamma\Lb \omega_2\Rb}\frac{ \Gamma\Lb \omega_3+n - k_1 -k_2\Rb}{\Gamma\Lb \omega_3\Rb}\intl\!\!\! d^2 r_i d^2 r'_i d^2b_i d^2b'_i\prod^n_{i=1} 
\frac{G_{\pom}\Lb z_i\Rb}{N_0}\gamma^{BA}\Lb r_i,r'_i,\delta b\Rb\frac{G_{\pom}\Lb z'_i\Rb}{N_0}
\eea
Using \eq{TU2} and \eq{DA21} we obtain:
   
      \bea \label{SA13}
S\Lb z'\Rb &=&\frac{1}{3!} S^3_A\Lb b \Rb C^3\sum^\infty_{n=0} \frac{\Lb - 1\Rb^n}{n!}\!\!\! \Lb \frac{G_{\pom}\Lb z'\Rb}{N_0}\Rb^n \intl^{\epsilon + i \infty}_{\epsilon - i \infty}\!\!\! \frac{ d \omega}{2\,\pi\,i}  e^{ \frac{ \bg^2 \kappa}{2} \omega^2 }\frac{ \Gamma\Lb \omega+n\Rb}{\Gamma\Lb \omega\Rb}    \prod^3_{i=1} \,\,
\!\!\intl^{\epsilon + i \infty}_{\epsilon - i \infty}\!\!\!  \frac{d \omega_i}{2\,\pi\,i}\frac{1}{\Gamma\Lb \omega_i\Rb}  
e^{ \frac{ \bg^2 \kappa}{2} \omega_i^2 }\,\intl^{\infty}_0 d t_i e^{- t_i} t_i^{\omega_i -1 }\nn\\
& &\underbrace{\sum_{k=1}^{n-1} \frac{n!}{k_1!  k_2!(n -k_1 -k_2)!}  t_1^{k_1} 
t_2^{k_2 } t_3^{ n - k_1 - k_2 }}_{t_1^n + t_2^n+t_3^n -(t_1+t_2)^n- (t_1+t_3)^n -(t_2+t_3)^n + (t_1+t_2+t_3)^n}  \eea
 
 After summation over $n$ we have
   \bea \label{SA14}
S\Lb z'\Rb &=&\frac{1}{3!} S^3_A\Lb b \Rb C^3
\!\!\!  \intl^{\epsilon + i \infty}_{\epsilon - i \infty}\!\!\! \frac{ d \omega}{2\,\pi\,i} 
 e^{ \frac{ \bg^2 \kappa}{2}\omega^2}
\prod^3_{i=1}\,\,\intl^{\epsilon + i \infty}_{\epsilon - i \infty}\!\!\!  \frac{d \omega_i}{2\,\pi\,i} e^{ \frac{ \bg^2 \kappa}{2}\omega_i^2 }
\frac{1}{\Gamma\Lb \omega_i\Rb}\exp\Lb -\omega\, \ln\Lb \frac{G_{\pom}\Lb z'\Rb}{N_0}\Rb\Rb\\
& &\intl^{\infty}_0\!\!\!d t_1 \!\!\intl^{\infty}_0 \!\!\!d t_2 \!\!\intl^{\infty}_0 \!\!\!d t_3 e^{ - t_1 - t_2 -t_3} 
\,\Bigg\{t_1^{-\omega} + t_2^{-\omega} +t_3^{-\omega}   - (t_1+t_2)^{-\omega}   - (t_1+t_3)^{-\omega} -(t_2+t_3)^{-\omega} + (t_1+t_2+t_3)^{-\omega} \Bigg\} \nn    \eea    
    
 Taking integrals over $t_i$ we obtain:
  \begin{subequations}    \bea 
S\Lb z'\Rb &=&\frac{1}{3!} S^3_A\Lb b \Rb C^3
\!\!\!  \intl^{\epsilon + i \infty}_{\epsilon - i \infty}\!\!\! \frac{ d \omega}{2\,\pi\,i} 
 e^{ \frac{ \bg^2 \kappa}{2}\omega^2}
\prod^3_{i=1}\intl^{\epsilon + i \infty}_{\epsilon _ i \infty}\!\!\!  \frac{d \omega_i}{2\,\pi\,i} e^{ \frac{ \bg^2 \kappa}{2}\omega_i^2 }
\exp\Lb -\omega\, \ln\Lb \frac{G_{\pom}\Lb z'\Rb}{N_0}\Rb\Rb\label{SA15a}\\
& &
\,\Bigg\{\underbrace{\sum^3_{i=1}\frac{ \Gamma\Lb \omega_i - \omega\Rb}{ \Gamma\Lb \omega_i \Rb} }_{\rm I}-\underbrace{\frac{\Gamma\Lb \omega_1+\omega_2 - \omega\Rb}{\Gamma\Lb \omega_1+\omega_2 \Rb}- \frac{\Gamma\Lb \omega_1+\omega_3 - \omega\Rb}{\Gamma\Lb \omega_1+\omega_3 \Rb}  - \frac{\Gamma\Lb \omega_2+\omega_3 - \omega\Rb}{\Gamma\Lb \omega_2+\omega_3 \Rb} }_{\rm II} + \underbrace{ \frac{\Gamma\Lb \omega_1+\omega_2+\omega_3 - \omega\Rb}{ \Gamma\Lb \omega_1+\omega_2+\omega_3 \Rb}
\Bigg\} }_{\rm III}\nn\\
&=&\frac{1}{3!} S^3_A\Lb b \Rb C^3\Bigg\{\underbrace{ 3\,C'  \exp\Lb - \frac{ z'^2}{4 \kappa}\Rb}_{\rm I} -3 \underbrace{  C'  \exp\Lb - \frac{ z'^2}{3 \kappa}\Rb}_{\rm II}\,+\,\underbrace{  C'  \exp\Lb - \frac{3 z'^2}{8 \kappa}\Rb}_{\rm III}\Bigg\}\label{SA15b}\\
&\xrightarrow{z \gg 1} &-\h S^3_A\Lb b \Rb  C'  \exp\Lb - \frac{ z'^2}{4 \kappa}\Rb\label{SA15c} \eea    
 \end{subequations}
 \eq{SA15b} is derived from \eq{SA15a} performing integration over $\omega$ closing contour on the poles of $\Gamma$ functions.

       ~

     ~  

     \begin{boldmath}
     \subsection{General case}
 \end{boldmath}   
   
         ~

         ~
        
        ~
        
       The scattering amplitude in  the general case of arbitrary $j$ can be written as follows:
         \bea      
S\Lb z\Rb &=&\frac{(-1)^{j+1}}{j!} S^j_A\Lb b \Rb C^j
\!\!\!  \intl^{\epsilon + i \infty}_{\epsilon - i \infty}\!\!\! \frac{ d \omega}{2\,\pi\,i} 
 e^{ \frac{ \bg^2 \kappa}{2}\omega^2}
\prod^j_{i=1}\intl^{\epsilon + i \infty}_{\epsilon _ i \infty}\!\!\!  \frac{d \omega_i}{2\,\pi\,i} e^{ \frac{ \bg^2 \kappa}{2}\omega_i^2 }
\exp\Lb -\omega\, \ln\Lb \frac{G_{\pom}\Lb z'\Rb}{N_0}\Rb\Rb\label{SA16}\\
& &
\sum^j_{j'=1}(-1)^{ j'+1}\frac{j!}{j'!\, (j - j')!}  \frac{ \Gamma\Lb \sum_{i=1}^{j'} \omega_i  \,-\,\omega\Rb}{\Gamma\Lb \sum_{i=1}^{j'}\omega_i \Rb} \nn
\eea
We close the contour of integration on the pole $\sum_{i=1}^{j'} \omega_i  \,-\,\omega =0$ reducing the term with $j'$ in $N_A\Lb z \Rb (N^{j'}_A\Lb z \Rb)$ to the form:
\beq \label{SA161}
S^{j'}\Lb z\Rb =\frac{(-1)^{j+1}}{j!} S^j_A\Lb b \Rb C^j
\!\!\!  \intl^{\epsilon + i \infty}_{\epsilon - i \infty}\!\!\! \frac{ d \omega}{2\,\pi\,i} 
\prod^j_{i=2}\intl^{\epsilon + i \infty}_{\epsilon _ i \infty}\!\!\!  \frac{d \omega_i}{2\,\pi\,i} e^{ \h\bg^2 \kappa\Lb\omega^2 +\Lb \omega - \sum_{i=2}^{j'}\omega_i\Rb^2 +\sum_{i=2}^{j'} \omega_i^2 \Rb}\exp\Lb -\omega\, \ln\Lb \frac{G_{\pom}\Lb z'\Rb}{N_0}\Rb\Rb
\eeq
Expecting that $\omega \propto \,\bg z$ is large we can take the integrals over $\omega$'s using the method of steepest descent . The equations for the saddle points are the following:
\beq \label{SA162}
(1)\,\,\, \bg^2 \kappa\Lb 2 \omega^{\mbox{\tiny SP}} - \sum_{i=2}^{j'}\omega^{\mbox{\tiny SP}} _i\Rb - \bg z\,=\,0\,;~~~~~(i) \,\,\, \omega^{\mbox{\tiny SP}}_i\,+\,\sum_{i=2}^{j'}\omega^{\mbox{\tiny SP}} _i\,-\,\omega \,=\,0\,;
\eeq
Solutions to these equations are $\omega^{\mbox{\tiny SP}}_i =\frac{1}{j'} \omega$ and $ \omega^{\mbox{\tiny SP}} \,\,=\,\,\frac{j'}{j'+1}\frac{z}{\bg \,\kappa} $. Using these solutions we obtain for the scattering amplitude:

  \begin{subequations}   
   \bea S\Lb z\Rb &=&  \frac{(-1)^{j+1}}{j!} S^j_A\Lb b \Rb C^j\,\sum^j_{j'=1}(-1)^{ j'+1} \frac{j!}{j'!\, (j - j')!}C'  \exp\Lb -\frac{j'}{j'+1} \frac{ z'^2}{2 \kappa}\Rb\label{SA17a}\\
&\xrightarrow{z\gg 1} &\,\,\frac{(-1)^{j+1}}{(j - 1)!} S^j_A\Lb b \Rb C^j\, \exp\Lb - \frac{ z'^2}{4 \kappa}\Rb\label{SA17b} \eea        
 \end{subequations}
 
   From \eq{SA17a}   one can see that the main contribution stems from $j'=1$ which corresponds to the configuration shown in \fig{2exmpl}-c, viz.: $j -1$ nucleons deliver one dipole each to the rapidity $Y_0$ .  The solution to BK equation leads to the asymptotic behaviour  of the last term in the sum of \eq{SA17a} $j' = j$ for large value of $j$.
   
   In principle we cannot sum \eq{SA17b} over $j$. However, assuming  that the smooth function is equal to $C^j$ as  it is shown in \eq{SA17b} we can obtain the scattering amplitude  in the fortm
   
   \beq \label{SA18}
   S\Lb z; b \Rb \,\,=\,\,C\,S_A\Lb b\Rb \exp\Lb - C\,S_A\Lb b \Rb\Rb   \exp\Lb - \frac{ z'^2}{4 \kappa}\Rb
   \eeq
    The   integral over $b$ can be taken  in two limited models:  $S_A$ has a Gaussian form or $S_A$ is a cylinder of radius $R_A$, The result is
    \bea \label{SA19}
 \mbox{Gaussian model:} &&  S\Lb z' \Rb \,\,=\,\,\pi \,R^2_A \Lb 1 - \exp\Lb - C\,\frac{A}{\pi R^2_A}\Rb \Rb   \exp\Lb - \frac{ z'^2}{4 \kappa}\Rb;\nn\\
  \mbox{Cylindrical  model:} &&  S\Lb z' \Rb \,\,=\,\,C\,A\exp\Lb -C \frac{A}{\pi R^2_A}\Rb\,\,    \exp\Lb - \frac{ z'^2}{4 \kappa}\Rb; \eea
    where $C$ has dimension of the cross section.

    Rewriting the expression in \eq{SA19} for the cylindrical model in the simplified form neglecting pre-exponential factors we have
    
    \beq \label{SA20}
    S\Lb z'\Rb \,\,=\,\,C \,A\,\exp\Lb -c \,A^{1/3}\Rb\,\,    \exp\Lb - \frac{ z'^2}{4 \kappa}\Rb;    
    \eeq

        Comparing \eq{SA20} with the BK amplitude at large $z$:
        \beq \label{ SA21} 
        S^{BK}\Lb z'\Rb\,=\,\pi R^2_A \,C \exp\Lb - \frac{ z'^2}{2 \kappa}\Rb        
        \eeq
        
        one can see that for $z' \,>\,2\, \sqrt{\kappa\,c}A^{1/6} $ the BK equation gives the scattering amplitude which is smaller than the one which stems from the large Pomeron loops.

        ~     
      ~

   ~

   ~
   
     \begin{boldmath}
     \section{Conclusions }
      \end{boldmath}

      
      In this paper we  summed the large BFKL Pomeron loops in the framework of  the BFKL Pomeron calculus for the scattering of dipole with nucleus target.The main result is that this scattering amplitude has the same energy dependence as the dipole-dipole amplitude. It means that the non-linear BK equation cannot describe the contribution of the large Pomeron loops and can only be used  in the limited range of energy for the dipole-nucleus scattering:
  $z' \,\leq\, 2\sqrt{\,\kappa\,c}A^{1/6} $.  From \eq{SA4} for $z'$ one can see that our statement about energy actually related to the kinematic region where $\bas \frac{\chi\Lb \bg\Rb}{\bg} \,Y \,\,\gg\,\,\xi_{r,r'}$. For scattering with nuclei $\xi_{r,r'} \,\approx\, \ln \Lb \frac{r^2\,r'^2}{R^4_A}\Rb\approx \ln \Lb \frac{R_N^2}{Q^2\,R^4_A}\Rb $, where $R_N $  is the nucleon radius, $Q^2$ is the photon virtuality in deep inelastc scattering and $R_A$ is the radius of a nucleus. Therefore, we can reach the asymptotic behaviour at sufficiently large values of $Y$:  $\bas \frac{\chi\Lb \bg\Rb}{\bg} \,Y\,\,>\,2\sqrt{\,\kappa\,c}A^{1/6} + \ln\Lb Q^2\,R^2_N\Rb +\frac{4}{3}\ln A \approx\,  18 +  \ln\Lb Q^2\,R^2_N\Rb$ for $A\sim 200$ and $c \sim 1$. Hence, practically we expect that we can trust BK equation for  accessible    energies.

 We wish to emphasize that the estimate for $z'$  were made using several assumptions which we cannot prove. First, it is assumed that the smooth function of $z$ for the interaction with $j$-nucleons is equal $C^J\Lb z\Rb $. Second, we used cylindrical model to evaluate the cross sections. 
     Third, we actually assumed that $ C^j\Lb z\Rb $ tends to constant at $z \to \infty$. Bearing this in mind we  are aware that our estimates are not very reliable.  The result, that does not depends on  additional assumptions, is that  dipole-nucleus scattering amplitude at ultrahigh energy behaves  as the amplitude of the dipole-dipole scattering.

      The result has been expected since it is shown in Refs.\cite{KLLL1,KLLL2}      
      that BK equation violates the s-channel unitarity.  The kind of surprise is the resulting scattering amplitude coincides    with the dipole-dipole scattering at high energy.

    We hope that this paper will contribute to the further study of the Pomeron calculus in QCD.

    ~

    ~

       {\bf Acknowledgements} 
     
   We thank our colleagues at Tel Aviv university  for
 discussions. Special thanks go A. Kovner and M. Lublinsky for stimulating and encouraging discussions on the subject of this paper. 
  This research was supported  by 
BSF grant 2022132. 

\appendix

~

~

~
    \begin{boldmath}

      \section{Useful formulae for $\rho^T_n\Lb r,b, \{r_i,b_i\}\Rb$  }     
    \end{boldmath}~


~
One can check that
\beq \label{DA3}
\frac{1}{k_i!} \frac{\Gamma\Lb \omega_i +k_i\Rb}{\Gamma\Lb \omega_i \Rb}\,=\,\frac{\sin\Lb \pi \omega_i\Rb}{\pi}\int^1_0 d t_i \Lb \frac{t_i}{1 - t_i }\Rb^{\omega_i} t_i^{k_i - 1}
\eeq
Plugging this equation in \eq{DA2} we have

 \bea\label{DA4}
  \rho^T_n\Lb r,b, \{r_i,b_i\}\Rb   \,\,&=&\,\,\prod^n_{i=1} \frac{G_{\pom}\Lb z_i\Rb}{N_0}
 \sum^{n}_{j=1}  \frac{1}{j!} S^j_A\Lb b\Rb\!\!\!\!\!\!\!\!\!\!\!\!\!\!\!\sum^{k_1=n - j+1,\dots\,k_j=n-j+1}_{k_1=1,\dots\,k_j=1}\!\!\!\!\!\!\!\!\!\!\!\!\!\!\!\!\!\!\!\!\delta_{\sum _{i=1}^j k_i = n}\nn\\
 &\times&  \prod^{j}_{i=1} 
C \,\intl^{\epsilon + i \infty}_{\epsilon - i \infty} \frac{d \omega_i}{2\,\pi\,i}\int^1_0 d t_i e^{ \frac{ 
\bg\,\kappa}{2}\,\omega_i^2}\frac{\sin\Lb \pi \omega_i\Rb}{\pi}\Lb\frac{t_i}{1 - t_i }\Rb^{\omega_i} t_i^{k_i - 1}
 \eea

 Replacing $\delta_{\sum _{i=1}^j k_i = n} $ by
 \beq \label{DA5}
 \delta_{\sum _{i=1}^j k_i = n} \,\,=\,\, \oint_C \mu^{ n - \sum^j_{i=1}k_i   -1} d \mu
 \eeq
 with contour  C being the circle around $\mu=0$, we obtain the following $ \rho^T_n\Lb r,b, \{r_i,b_i\}\Rb $ after summing over $k_i$:
 
   \bea\label{DA4}
  \rho^T_n\Lb r,b, \{r_i,b_i\}\Rb   \,\,&=&\,\,\prod^n_{i=1} \frac{G_{\pom}\Lb z_i\Rb}{N_0}
 \sum^{n}_{j=1}  \frac{1}{j!} S^j_A\Lb b\Rb\oint_C \mu^{ n   -1} d \mu
\nn\\
 &\times&  \prod^{j}_{i=1} 
C \,\intl^{\epsilon + i \infty}_{\epsilon - i \infty} \frac{d \omega_i}{2\,\pi\,i}\int^1_0 d t_i e^{ \frac{ 
\bg\,\kappa}{2}\,\omega_i^2}\frac{\sin\Lb \pi \omega_i\Rb}{\pi}\Lb\frac{t_i}{1 - t_i }\Rb^{\omega_i} \frac{1}{\mu} \frac{ 1 - \Lb \frac{t_i}{\mu}\Rb^{n - j+1}}{1 - \Lb \frac{t_i}{\mu}\Rb}  \eea
  Taking integrals over $t_i$ we have:
  
     \bea\label{DA41}
  \rho^T_n\Lb r,b, \{r_i,b_i\}\Rb   \,\,&=&\,\,\prod^n_{i=1} \frac{G_{\pom}\Lb z_i\Rb}{N_0}
 \sum^{n}_{j=1}  \frac{1}{j!} S^j_A\Lb b\Rb\oint_C \mu^{ n   -1} d \mu
  \prod^{j}_{i=1} 
C \,\intl^{\epsilon + i \infty}_{\epsilon - i \infty} \frac{d \omega_i}{2\,\pi\,i}e^{ \frac{ 
\bg\,\kappa}{2}\,\omega_i^2}\Bigg\{ \left((\mu -1)^{-\omega } \mu ^{\omega }-1\right) \nn\\
&-&\sin\Lb \pi \omega_i\Rb\Gamma (1-\omega ) \left(\frac{1}{\mu }\right)^{-j+n+2}\!\!\!\!\!\!\!\!\!\!\!\!\!\! \Gamma (-j+n+\omega +2) \, _2\tilde{F}_1\left(1,-j+n+\omega +2;-j+n+3;\frac{1}{\mu }\right)\Bigg\}\eea

\end{document}